\documentstyle[epsfig]{aipproc}
\def\Journal#1#2#3#4{{#1} {\bf #2}, #3 (#4)}

\def\NPB{{\em Nucl.~Phys.}~B}
\def\PLB{{\em Phys.~Lett.}~B}
\def\PRL{\em Phys.~Rev.~Lett.}
\def\PRD{{\em Phys.~Rev.}~D}

\def\be{\begin{equation}}
\def\ee{\end{equation}}
\def\bea{\begin{eqnarray}}
\def\eea{\end{eqnarray}}
\def\ba{\begin{array}}
\def\ea{\end{array}}
\def\nn{\nonumber}
\def\simge{\mathrel{%
   \rlap{\raise 0.511ex \hbox{$>$}}{\lower 0.511ex \hbox{$\sim$}}}}
\def\simle{\mathrel{
   \rlap{\raise 0.511ex \hbox{$<$}}{\lower 0.511ex \hbox{$\sim$}}}}
 
\def\slashchar#1{\setbox0=\hbox{$#1$}           
   \dimen0=\wd0                                 
   \setbox1=\hbox{/} \dimen1=\wd1               
   \ifdim\dimen0>\dimen1                        
      \rlap{\hbox to \dimen0{\hfil/\hfil}}      
      #1                                        
   \else                                        
      \rlap{\hbox to \dimen1{\hfil$#1$\hfil}}   
      /                                         
   \fi}                                         %
\def\ts{\thinspace}

\def\ra{\rightarrow}

\def\ol{\bar}

\def\CO{{\cal O}}

\def\atc{\alpha_{TC}}

\def\Few{F_\pi}
\def\Ntc{N_{TC}}

\def\getc{g_{ETC}}

\def\kslash{\raise.15ex\hbox{/}\kern-.57em k}
\def\LTC{\Lambda_{TC}}

\def\METC{M_{ETC}}

\def\condtc{\langle \ol T T \rangle_{TC}}
\def\condetc{\langle \ol T T \rangle_{ETC}}
\def\tro{\rho_{T}}

\def\troz{\rho_{T}^0}
\def\tropm{\rho_{T}^\pm}

\def\tom{\omega_T}
\def\tpi{\pi_T}

\def\mev{{\rm MeV}}
\def\gev{{\rm GeV}}
\def\tev{{\rm TeV}}

\def\ifb{{\rm fb}^{-1}}
\def\half{{\textstyle{ { 1\over { 2 } }}}}

\def\eighth{{\textstyle{{1\over {8}}}}}

\def\nin{\noindent}

\begin{document}

\title{
\vskip -15mm
\begin{flushright}
\vskip -15mm
{\small BUHEP-99-30\\
hep-ph/9912526\\}
\vskip 5mm
\end{flushright}
{\Large{\bf{Technicolor Signatures at the \\ High Energy Muon Collider}}\\
}}


\author{Kenneth Lane\thanks{lane@buphyc.bu.edu}$^,$\thanks{Talk delivered at
     the workshop ``Studies on Colliders and Collider Physics at the Highest
     Energies: Muon Colliders at 10 TeV to 100
     TeV'', Montauk, Long Island, NY, 27~September--1~October 1999.}}
\address{Department of Physics\\
Boston University\\
590 Commonwealth Avenue\\
Boston, Massachussets 02215\\}

\maketitle

\begin{abstract}
I discuss high mass signatures of technicolor that would be observable at a
very high energy muon collider. Most intriguing is the spectrum of spin--one
technihadrons, $\tro$, $\tom$ and $A_{1T}$, which may extend to 100~TeV and
beyond in a walking technicolor theory.

\end{abstract}

\section*{1. Introduction}

It is a real pleasure to talk at a workshop in which the theorists are
down--to--earth participants and the machine physicists are wild--eyed
dreamers. Here is an e-mail exchange between between my session organizer and
me:

\begin{itemize} 

\item Joe --

I just realized that the workshop title refers to muon colliders at 10-100
TeV~(!). I don't have a hell of a lot in the way of TC signals at those
energies. How seriously should I take that energy range as a charge??

  Ken

\item You can completely ignore the 10 TeV stuff - that is
for the accelerator people (i.e. what is the highest
energy muon collider one could ever have any hope of
building).

--Joe

\end{itemize}

\nin Accordingly, I prepared a talk that discusses TC signatures at 1--4~TeV:
The technivector mesons $\tro$ and $\tom$ of the minimal, one--doublet TC
model~\cite{tc}; the $Z'$ and higher--dimensional electroweak singlet
technifermions of topcolor--assisted technicolor~\cite{tctwohill}; and the
electroweak--$SU(2)$ singlet fermions of the top seesaw model~\cite{seesaw}.

In the course of this, however, I recalled an old idea that would give the
HEMC physics to do all the way from 1~TeV to 100~TeV. This has to do with the
fact that walking technicolor~\cite{wtc}, an essential ingredient of any
viable TC model, implies that the spectrum of technivector mesons cannot be
QCD--like~\cite{tasi,ichep94,eduardo}. It must extend in some sense to
100~TeV and beyond. This idea is so intriguing that I will emphasize it
here. I hope someone will be able to decide whether it makes sense.

The rest of this paper is organized as follows: Section~2 presents a summary
of the dynamical approach to electroweak and flavor symmetry breaking:
technicolor~\cite{tc}, extended technicolor (ETC)~\cite{etcsd,etceekl}, and
all that. This scenario's signatures at the HEMC are discussed in Section~3,
with emphasis on the technivector spectrum in walking technicolor models.

\section*{2. Overview of Technicolor} 

Technicolor---a strong interaction of fermions and gauge bosons at the scale
$\LTC \sim 1\,\tev$---induces the breakdown of electroweak symmetry to
electromagnetism {\em without} elementary scalar bosons~\cite{tc}.
Technicolor has a strong precedent in QCD. There, the chiral symmetry of
massless quarks is spontaneously broken by strong QCD interactions, resulting
in the appearance of massless Goldstone bosons, $\pi$, $K$,
$\eta$.~\footnote{The hard masses of quarks explicitly break chiral symmetry
and give mass to $\pi$, $K$, $\eta$, which are then referred to as
pseudo-Goldstone bosons.} In fact, if there were no Higgs bosons, this chiral
symmetry breaking would itself cause the breakdown of $SU(2) \otimes U(1)$ to
electromagnetism. Furthermore, the $W$ and $Z$ masses would be given by
$M_W^2 = M_Z^2 \cos^2\theta_W = \eighth g^2 N_F f_\pi^2$, where $g$ is the
weak $SU(2)$ coupling, and $N_F$ the number of massless quark flavors. Alas,
the pion decay constant $f_\pi$ is only $93\,\mev$ and the $W$ and $Z$ three
orders of magnitude too light.

In its simplest form, technicolor is a scaled up version of QCD, with
massless technifermions whose chiral symmetry is spontaneously broken at
$\LTC$. If left and right-handed technifermions are assigned to weak $SU(2)$
doublets and singlets, respectively, then $M_W = M_Z \cos\theta_W = \half g
\Few$, where $\Few = 246\,\gev$ is the weak {\em technipion} decay
constant.~\footnote{In the minimal model with one doublet $(U,D)$ of
technifermions, there are just three technipions. They are the linear
combinations of massless Goldstone bosons that become, via the Higgs
mechanism, the longitudinal components $W_L^\pm$ and $Z_L^0$ of the weak
gauge bosons. In non-minimal technicolor, the technipions include the
longitudinal weak bosons as well as additional Goldstone bosons associated
with spontaneous technifermion chiral symmetry breaking. The latter must and
do acquire mass---from the extended technicolor interactions discussed
below.}

In the standard model and its extensions, the masses of quarks and leptons
are produced by their Yukawa couplings to the Higgs bosons---couplings of
arbitrary magnitude and phase that are put in by hand. This option is not
available in technicolor because there are no elementary scalars. Instead,
this explicit breaking of quark and lepton chiral symmetries must arise from
{\it gauge interactions alone}. The most economical approach employs extended
technicolor~\cite{etcsd,etceekl}. In its proper formulation~\cite{etceekl},
the ETC gauge group contains technicolor, color, and flavor as subgroups and
there are very stringent restrictions on the representations to which
technifermions, quarks, and leptons belong: Specifically, they must be
combined into the same few large representations of ETC. Otherwise, unbroken
chiral symmetries lead to axion--like particles. Quark and lepton hard masses
are generated by their coupling (with strength $\getc$) to technifermions via
ETC gauge bosons of generic mass $\METC$:
\be\label{eq:qmass}
m_q(\METC) \simeq m_\ell(\METC)  \simeq {\getc^2 \over
{\METC^2}} \condetc \ts,
\ee
where $\condetc$ and $m_{q,\ell}(\METC)$ are, respectively, the technifermion
condensate and quark and lepton masses renormalized at the scale $\METC$.

Technicolor is an asymptotically free gauge interaction. If it is like QCD,
with its running coupling $\atc$ rapidly becoming small above its
characteristic scale $\LTC \sim 1\,\tev$, then $\condetc \simeq \condtc
\simeq \LTC^3$. To obtain quark masses of a few~GeV thus requires
$\METC/\getc \simle 30\,\tev$. This is excluded: Extended technicolor boson
exchanges also generate four-quark interactions which, generically, include
$|\Delta S| = 2$ and $|\Delta B| = 2$ operators. For these not to conflict
with $K^0$-$\ol K^0$ and $B_d^0$-$\ol B_d^0$ mixing measurements,
$\METC/\getc$ must exceed {\it several hundred} TeV~\cite{etceekl}. This
implies quark and lepton masses no larger than a few MeV, and technipion
masses no more than a few~GeV.

Because of this conflict between constraints on flavor-changing neutral
currents and the magnitude of ETC-generated quark, lepton and technipion
masses, classical QCD--like technicolor was superseded long ago by
``walking'' technicolor~\cite{wtc}. Here, the strong technicolor coupling
$\atc$ runs very slowly, or walks, for a large range of momenta, possibly all
the way up to the ETC scale of {\it several hundred} TeV. The slowly-running
coupling enhances $\condetc/\condtc$ by almost a factor of
$\METC/\LTC$. This, in turn, allows quark and lepton masses as large as a
few~GeV and $M_{\tpi} \simge 100\,\gev$ to be generated from ETC interactions
at $\METC = \CO(100\,\tev)$.

In almost all respects, walking technicolor models are very different from
QCD with a few fundamental $SU(3)$ representations. One example is that
integrals of weak-current spectral functions and their moments converge much
more slowly than they do in QCD. The consequence of this for the HEMC will be
discussed in Section~3. Meanwhile, this and other calculational tools based
on naive extrapolation from QCD and on large-$\Ntc$ arguments are suspect. It
is not yet possible to predict with confidence the influence of technicolor
degrees of freedom on precisely-measured electroweak quantities---the $S,T,U$
parameters to name a frequently discussed example~\cite{pettests}.

Another major development in technicolor was motivated by the discovery of
the top quark at Fermilab~\cite{toprefs}. Theorists have concluded that ETC
models cannot explain the top quark's large mass without running afoul of
either experimental constraints from the $\rho$ parameter and the $Z \ra \ol
b b$ decay rate~\cite{zbbth}---the ETC mass must be about 1~TeV to produce
$m_t = 175\,\gev$; see Eq.~(\ref{eq:qmass})---or of cherished notions of
naturalness---$\METC$ may be higher, but the coupling $\getc$ then must be
fine-tuned near to a critical value. This state of affairs led to the
proposal of ``topcolor-assisted technicolor'' (TC2)~\cite{tctwohill}.

In TC2, as in many top-condensate models of electroweak symmetry
breaking~\cite{topcondref}, almost all of the top quark mass arises from a
new strong ``topcolor'' interaction~\cite{topcref}. To maintain electroweak
symmetry between (left-handed) top and bottom quarks and yet not generate
$m_b \simeq m_t$, the topcolor gauge group under which $(t,b)$ transform is
usually taken to be $SU(3)\otimes U(1)$. The $U(1)$ provides the difference
that causes only top quarks to condense. Then, in order that topcolor
interactions be natural---i.e., that their energy scale not be far above
$m_t$---without introducing large weak isospin violation, it is necessary
that electroweak symmetry breaking is still due mostly to technicolor
interactions~\cite{tctwohill}.

Extended technicolor interactions are still needed in TC2 models to generate
the masses of light quarks and the bottom quark, to contribute a few~GeV to
$m_t$,~\footnote{Massless Goldstone ``top-pions'' arise from top-quark
condensation. This ETC contribution to $m_t$ is needed to give them a mass in
the range of 150--250~GeV.} and to give mass to technipions. The scale of ETC
interactions still must be hundreds of~TeV to suppress flavor-changing
neutral currents and, so, the technicolor coupling still must walk.  In TC2
there is no need for large technifermion isospin splitting associated with
the top-bottom mass difference. Thus, for example, $\tom$ and $\tro$ partners
are nearly degenerate $\ol UU \pm \ol DD$ states.

Another, more recent, variant of topcolor models is the ``top seesaw''
mechanism~\cite{seesaw}. Its motivation is to realize the original
top--condensate idea of the Higgs boson as a fermion--antifermion bound
state. This failed for the top quark because it turned out to be too light!
In top seesaw models, an electroweak singlet fermion $F$ acquires a dynamical
mass of {\it several} TeV. Through mixing of $F$ with the top quark, it gives
the latter a much smaller mass (the seesaw) and the scalar $\ol F F$ bound
state acquires a component with an electroweak symmetry breaking vacuum
expectation value.

This completes our brief summary of technicolor. We turn now to the
technicolor signatures for which a high energy muon collider is
well--suited.

\section*{3. Technicolor Signatures at the HEMC}

The principal signals of technicolor are discussed in a number of places~
\cite{lowsigs}. Most of them are accessible at low energies---at the Tevatron
in Run~II, certainly at the LHC, and, possibly, even at LEP.  In the minimal
technicolor model, with just one technifermion doublet, the only prominent
signals in a TeV--scale collider are modest enhancements in
longitudinally-polarized weak boson production. These are the $s$--channel
color--singlet technirho resonances near 1.5--2~TeV: $\troz \ra W_L^+W_L^-$
and $\tropm \ra W_L^\pm Z_L^0$. The $\CO(\alpha^2)$ cross sections of these
processes are quite small at such masses. This and the difficulty of
reconstructing weak-boson pairs with reasonable efficiency make observing
these enhancements a challenge. These states would be more easily seen in a
lepton collider---if one can be built with $\sqrt{s} = 1.5$--$2\,\tev$ at an
affordable cost. Nonminimal technicolor models are much more accessible in a
hadron collider because they have a rich spectrum of lower mass technirho
vector mesons and technipion states into which they may decay.

If technicolor is the basis for electroweak symmetry breaking, it will have
been discovered once the LHC has acquired and analyzed $10\,\ifb$ of
data. The question we address here is what the HEMC can do to add to our
understanding of this new dynamics.

\subsection*{3.1 The Technivector Spectrum of Walking Technicolor}

The slow decrease with energy of the coupling $\atc$ in walking technicolor
means that the $\mu^+\mu^-$ cross section approaches asymptotia only near the
extended technicolor scale, probably even above the reach of the HEMC. This
is most directly seen by considering the integrals in Weinberg's spectral
function sum rules for the weak--isospin vector and axial vector
currents~\cite{sfsr}. These sum rules are
\bea\label{eq:spectral}
& &\int_0^\infty ds \left [\rho_V(s) - \rho_A(s) \right] = F_\pi^2 \nn\\ 
& & \int_0^\infty ds \ts s\ts\left[\rho_V(s) - \rho_A(s) \right] = 0
\ts,
\eea
where $F_\pi = 246\,\gev$. Here, the spectral functions $\rho_V$ and $\rho_A$
are analogs for the weak--isospin currents of the ratio of cross sections,
$R(s) = \sigma(e^+e^- \ra {\rm hadrons}) / \sigma(e^+e^- \ra \mu^+
\mu^+)$. In QCD, the sum rules corresponding to Eq.~(\ref{eq:spectral}) are
saturated by the lowest lying spin--one resonances, $\rho$ and $A_1$, and the
sum rules converge rapidly above the $A_1$ mass. Similarly, in technicolor
without a walking coupling, the sum rules would be saturated by the lowest
$\tro$ and $A_{1T}$ and the difference $\rho_V - \rho_A \sim 1/s^3$ for $s
\simge M^2_{A_{1T}} \sim 1\,\tev^2$. In walking technicolor, the slow running
of $\atc(s)$ implies that $\rho_V - \rho_A \sim 1/s^2$ below $s \sim
M_{ETC}^2$ and $1/s^3$ above.  Thus, {\it the spectral functions cannot be
saturated by a single pair of low--lying resonances.} Either there must be a
tower of resonances above $\tro$ and $A_{1T}$, all of which contribute
significantly to the spectral integrals (see Ref.~\cite{tasi,ichep94}; also
Ref.~\cite{eduardo} for an explicit attempt to realize this), or the spectral
functions are smooth but anomalously slowly decreasing up to $M_{ETC}$. The
same alternative applies to the $\mu^+\mu^-$ cross section. Moreover, the
isoscalar state $\tom$ and its excitations appear there. Thus, exploration of
the 1--100~TeV region of $\mu^+\mu^-$ annihilation is bound to reveal crucial
information on the dynamics of a walking gauge theory, dynamics on which we
theorists can only speculate.

In the minimal one--doublet model of technicolor, it has always been assumed
that the lowest lying $\tro$, $\tom$, and $A_{1T}$ decay mainly into two and
three longitudinally--polarized weak bosons, $W^\pm_L$ and $Z^0_L$. In the
minimal model, however, $M_{\rho_T} \sim M_{A_{1T}} = 1$--$2\,\tev$, and this
is so far above $2M_W$ that it is possible that decay modes with more than
two or three weak bosons are important if not dominant.~\footnote{The QCD
$2\ts ^3S_1$ state $\rho'(1700)$ decays predominantly to four, not two pions,
presumably because the two--pion mode is suppressed by an exponential form
factor and/or a node in the decay amplitude.} Thus, in the minimal walking
technicolor model, there may be a tower of vector and axial vector mesons in
the $s$--channel of $\mu^+\mu^-$ annihilation which decay to many $W$ and $Z$
bosons. It is an open question how narrow and discernible these resonances
will be.

In nonminimal models, the spectrum of technihadrons is quite rich and the
scale of their masses is lower (roughly as the square root of the number of
technifermion doublets). There are technipions $\tpi$ as well as weak bosons
for the $\tro$, $\tom$, and $A_{1T}$ to decay into. These $\tpi$ may be color
singlets and, if colored technifermions exist, octets and triplets
(``leptoquarks''). Technipions are expected to have masses in the range
100---500~GeV and to decay into the heaviest fermion pairs allowed. The large
value of $\condetc/\condtc$ in walking technicolor significantly enhances
technipion masses. Thus, for example, $\rho_T \ra \tpi\tpi$ decay channels
may be closed for the lowest--lying state. Instead, $\tro \ra W_L W_L$, $W_L
\tpi$, and $\gamma \tpi$~\cite{lowsigs}. The excited states should be able to
decay into pairs of technipions. The $\tro$, $\tom$, and $A_{1T}$ that lie
above multi--$\tpi$ threshold are likely to be wider than their counterparts
in the minimal model. Still, the structure of $\mu^+\mu^-$ annihilation up to
100~TeV will provide valuable insight to walking gauge dynamics.

\subsection*{3.2 Topcolor--Technicolor Signals}

As I said above, topcolor--assisted technicolor generally employs an extra
``hypercharge'' $U(1)$ to help induce a large condensate for the top, but not
the bottom quark. This additional $U(1)$ is broken, leading to a $Z'$ boson
which is strongly coupled to at least the third generation. In the models of
Ref.~\cite{tctwoklee}, it is strongly coupled to all fermions. Some of the
lower energy phenomenology of this $Z'$ was studied in
Refs.~\cite{bonini,rador}. Its nominal mass, in the range 1--4~TeV, and
potentially strong coupling to muons make it a target of opportunity for the
HEMC.~\footnote{Top seesaw models also have an extra $U(1)$ gauge symmetry,
broken spontaneously. There, the $Z'$ boson mass is expected to be roughly
5~TeV.} Unfortunately, its strong couplings and many decay channels to
ordinary fermions and technifermions may also make the $Z'$ so broad that it
is difficult discover and study in any collider.

An intriguing feature of this $Z'$ is that it must acquire its mass from
condensation of a technifermion $\psi$~\cite{tctwoklee}. The $Z'$ mass of
several TeV implies that the $\psi$--fermion's mass is 1--2~TeV. Thus, $\psi$
must transform according to a higher--than--fundamental representation of the
technicolor gauge group. In order that its condensation not break electroweak
$SU(2) \otimes U(1)$, $\psi$ must either be a singlet or transform
vectorially under this symmetry. The obvious way to access it is via $Z' \ra
\ol \psi \psi$ in the $s$--channel of the HEMC. The phenomenology of these
higher representation technifermions has not been studied in detail. One
crucial question is whether $\psi$ is stable. If not, how does it decay? If
it is, what are the cosmological consequences?

Finally, there is the $SU(2)$ singlet, charge--2/3 quark $F$ of top seesaw
models. This fermion also has a mass of several TeV and may be pair produced
via $\gamma,Z,Z'$ at the HEMC. It decays by virtue of its mixing with the top
quark as $F \ra t \ra Wb$, a striking signature indeed.

\section*{4. Conclusions and Acknowledgements} 

The HEMC technicolor signatures that I have presented here are, quite
obviously, at a primitive stage of development. I think all of them deserve
further thought because they bear directly on unfamiliar dynamics such as
walking technicolor and strongly--coupled topcolor. Corresponding
uncertainties face the design of the HEMC. Again, the particle theorists and
the accelerator theorists are in the same boat. The need to go on to higher
energies remains and it always will. This was said very well by an Amherst
poet long ago:

\medskip

\begin{verse}
``Faith'' is a fine invention \\
When Gentlemen can see ---  

But  {\it Microscopes}  are prudent \\
In an Emergency.

\vskip0.15truein

\hskip0.5truein --- {\sl Emily Dickinson, 1860}

\end{verse}

I thank the organizers, especially Bruce King and Joe Lykken for inviting me
to this stimulating workshop and for the wonderful opportunities to explore
Montauk and Block Island. Kathleen Tuohy ran a perfect workshop and I send
her my gratitude. I am grateful to my fellow participants in the joint
Physics and Detector Working Group. They provided the mental stimulation that
led to my contribution. I am also indebted to Sekhar Chivukula for
discussions about top seesaw models and for reading this manuscript. This
research was supported in part by the Department of Energy under
Grant~No.~DE--FG02--91ER40676.

\end{document}